\documentclass[aps,prd,%twocolumn,
eqsecnum,showpacs,nofootinbib,preprintnumbers]{revtex4}

\usepackage{amsmath,amsfonts,amssymb,color,graphicx,latexsym,theorem,mathrsfs
}

\newcommand{\ma}[1]{\mbox{$\mathcal{#1}$}}
\newcommand{\mas}[1]{\mbox{$\mathscr{#1}$}}

\newcommand{\D}{{\rm d}}

\newcommand{\ti}{\tilde}
\newcommand{\we}{\wedge}

\begin{document}
 \preprint{YITP-21-21}

%%%%%%%%%%%%   Title   %%%%%%%%%%%%%
\title{
An alternative to the Simon tensor
}

%%%%%%%%%%%%   Authors   %%%%%%%%%%%%%
\author{Masato Nozawa}
\email{
masato.nozawa@oit.ac.jp
}
%masato.nozawa@yukawa.kyoto-u.ac.jp

%%%%%%%%%%%%   Address   %%%%%%%%%%%%%
\address{ 
Department of General Education, Faculty of Engineering, Osaka Institute of Technology,
5-16-1, Omiya, Asahi-ku, Osaka, Osaka 535-8585, Japan.
%Center for Gravitational Physics, Yukawa Institute for Theoretical Physics, 
%Kyoto University, Kyoto 606-8502, Japan,\\
%General Education, Faculty of Engineering, 
%Osaka Institute of Technology, Osaka, Osaka 535-8585, Japan
}

\date{\today}

%%%%%%%%%%%%%%%%%%%%%%%%%%%%%%%
%                                                                                                %
%                                      Abstract                                          %
%                                                                                                %
%%%%%%%%%%%%%%%%%%%%%%%%%%%%%%%
\begin{abstract} 
The Simon tensor gives rise to a local characterization of the Kerr-NUT family in the stationary class of vacuum spacetimes.  We find that a symmetric and traceless tensor in the quotient space of the stationary Killing trajectory offers a useful alternative to the Simon tensor. Our tensor is distinct from the spatial dual of the Simon tensor and 
illustrates the geometric property of the three dimensional quotient space more manifest.  The reconstruction procedure of the metric for which the generalized Simon tensor vanishes is spelled out in detail. 
We give a four dimensional description of this tensor in terms of the Coulomb part of the imaginary selfdual Weyl tensor, which corresponds to the generalization of the three-index tensor defined by Mars. This allows us to establish a new and simple criterion for the Kerr-NUT family: the gradient of the Ernst potential becomes the non-null eigenvector of the Coulomb part of the imaginary selfdual Weyl tensor. We also discuss the ${\rm SU}(1,2)$ covariant extension of the obstruction tensor into the Einstein-Maxwell system 
as an intrinsic characterization of the Kerr-Newman-NUT family. 
\end{abstract}

%\pacs{} 

\maketitle

%%%%%%%%%%%%%%%%%%%%%%%%%%%%%%%
%                                                                                                %
%                                   Introduction                                         %
%                                                                                                %
%%%%%%%%%%%%%%%%%%%%%%%%%%%%%%%

\section{Introduction}

During the last couple of years, we have witnessed a considerable advance in our ability to 
access black holes in our universe. With the advent of gravitational wave astronomy 
\cite{Abbott:2016blz} and the direct detection of black hole shadow with unprecedented precision \cite{Akiyama:2019cqa},  the Kerr solution \cite{Kerr:1963ud}--describing a stationary and axisymmetric rotating black hole in vacuum spacetimes--has been a rekindled subject of intensive research. 
First and foremost, central to the analysis of black holes in our universe is the uniqueness theorem \cite{Carter:1971zc,Robinson:1975bv,Mazur:1982db}, which allows us to center exclusively on the Kerr solution, provided that the spacetime eventually settles down to an equilibrium final state. On top of this astrophysical significance, the Kerr solution provides an interesting  and valuable arena for the (pseudo-)Riemannian geometry.  This topic is considerably vast and has been tackled by 
numerous authors.

The first noticeable earmark of the Kerr metric is that there exist two distinct shear-free null  geodesic congruences.  Due to the theorem provided by Goldberg and Sachs \cite{GS}, these optical aspects are closely tied to the algebraic property of the Weyl tensor. Working in the Newman-Penrose formalism \cite{Newman:1961qr}, one finds a preferred frame for which Einstein's equations are simplified, allowing one to obtain a number of algebraically special solutions of physical significance in a closed form~\cite{Kinnersley:1969zza}. 
This property is the key for the original discovery of the Kerr solution \cite{Kerr:1963ud}. 

Another distinctive feature of the Kerr solution is that it admits a nondegenerate Killing-Yano tensor \cite{Yano}. We refer the readers to \cite{Yasui:2011pr} for a recent comprehensive review.  The Killing-Yano tensor is an anti-symmetric generalization of the Killing vector and is viewed as a ``square-root'' of the Killing tensor, the latter of which enables one to get the constant of geodesic motion \cite{Carter:1968ks,Walker:1970un}. 
By dint of the integrability conditions, the existence of the Killing-Yano tensor demands the algebraically special nature of curvature. Additional intriguing properties emerge when 
Wick-rotated to the Riemannian signature, for which the Euclidean Kerr solution admits an integrable complex structure \cite{Klemm:2013eca} and a conformal K\"ahler structure \cite{Nozawa:2015qea}. These properties are combined with the Killing-Yano tensor to generate the enhanced Euclidean supersymmetry
\cite{Nozawa:2015qea,Nozawa:2017yfl} when the Weyl tensor is self-dual. 

The present paper specializes to yet another characterization of the Kerr solution provided by 
Simon \cite{Simon}. The three-index Simon tensor $S_{abc}$ is defined on the manifold of trajectory for the stationary Killing field and vanishes for the Kerr(-NUT) solution. 
Its construction is based on the fact that it reduces in the static case to the Cotton tensor, which measures the obstruction to the conformal flatness in three space and thereby describes the derivation from the Schwarzschild solution in the vacuum case \cite{israel,MRS,robinson,bunting}. 
In the  paper \cite{Simon}, 
it has been advocated that the local portion of asymptotically flat vacuum spacetime admitting the vanishing Simon tensor is isometrically embedded into the Kerr spacetime. 
Subsequently, Mars has given the elegant spacetime description of the Simon tensor in terms of the imaginary self-dual complexified Weyl tensor and its eigen-twoform \cite{Mars:1999yn,Mars:2000gb}.  Some pertinent topics regarding these tensors have been studied in \cite{Bini:2001ke,Bini:2004qf,Some:2014kfa,Beyer:2017ghs}.

In this paper, we would like to revisit the issue of local characterization of the Kerr solution from the perspective of 
Simon \cite{Simon}. 
As a matter of first priority, the geometric interpretation of the Simon tensor has remained puzzling.
To the best of our knowledge, the utmost available expertise about the Simon tensor is merely a complex generalization of the Cotton tensor. It therefore appears imperative to appreciate at a deeper level the geometric origin of the Simon tensor. Our second impetus emerges from the static case. In our recent paper \cite{Nozawa:2018kfk}, we have found a novel tensor field $H_{ab}$, which deserves a global characterization of the Schwarzschild metric (see (\ref{HabSch}) for definition).  Accordingly, this tensor $H_{ab}$ supersedes the Cotton tensor in the static vacuum spacetime. The tenor $H_{ab}$ is more user-friendly than the three-index Cotton  tensor and has been exploited to derive a noteworthy equation of divergence type, which is of essential help in proving the uniqueness theorem of static black holes. 
It is therefore natural to seek a similar simplification in the stationary case. This would yield insights into the geometric structure of the stationary quotient space and 
would be likely to alleviate a lot of anguish in deriving nontrivial equations of divergence type. 

This paper is intended to provide a useful alternative to the Simon tensor. In common with the static counterpart, our proposed tensorial field $N_{ab}$ is symmetric and tracefree, but is complex (see (\ref{Hab}) for definition). It turns out that the Simon tensor $S_{abc}$ is completely represented in terms of our tensor $N_{ab}$. The vanishing of the tensor  $N_{ab}$ enforces a tight restriction to the geometry which is compatible with equations of motion. 
As demonstrated by Perjes \cite{Perjes} (see also \cite{Krisch}), some degenerate cases of $S_{abc}=0$ give rise to solutions whose explicit metric forms are indeterminate. To the contrary, the metrics with $N_{ab}=0$ are constrained more severely including the degenerate cases. 
Likewise, one must fix the integration constant arising from $S_{abc}=0$ by boundary conditions to exclude the ``topological version'' of the Kerr-NUT solution, whereas this extra task is unnecessary in our case. It therefore seems fair to say that our tensor field $N_{ab}$ deserves an appropriate local characterization 
of the Kerr-NUT family.  
Another applicability of our tensor is that it can be spacetime covariant, {\it mutatis
mutandis},  in a simple way. We uncover that for the Kerr-NUT solution the derivative of the Ernst potential becomes a non-null eigenvector of the Coulomb part of the imaginary selfdual Weyl tensor. 
Furthermore, the extension of $N_{ab}$ into the electrovacuum case is fairly straightforward without any drastic jump in logic, while the electrovacuum Simon tensor in \cite{Bini:2004qf} has been deduced from the Mars tensor. See also \cite{Wong:2008zb} for the detailed supplementary conditions to be imposed for the spacetime characterization of the Kerr-Newman solution.

The remainder of the current paper is constructed as follows. 
In the ensuing section, we review the argument of Simon for the stationary vacuum spacetimes and discuss a 
static prototype for generalizing the Simon tensor. Our proposed alternative to the Simon tensor is given in  section \ref{sec:Simon}. A systematic classification for the local metrics admitting $N_{ab}=0$ is implemented. 
 We also give a spacetime description of this tensor. 
Section \ref{sec:Maxwell} discusses the extension to the Einstein-Maxwell system. 
Our conclusion is summed up in section \ref{sec:summary}. 

We use Greek letters $\mu, \nu,...$ for spacetime indices and Latin letters 
$a, b,...$ for indices on the quotient space of the stationary Killing field. 
We adopt $c=G=1$ units throughout the papaer. 

\section{Preliminary}

\subsection{Vacuum spacetime with a stationary Killing vector}

Let us consider the vacuum solution to Einstein's field equations $R_{\mu\nu}=0$
endowed with a stationary Killing vector field $\xi ^\mu$. 
In an adapted coordinate system  $x^\mu=(t, x^a)$ with $\xi =\partial/\partial t$, 
the metric can be written locally as
\begin{align}
\label{metric}
\D s^2=-f(\D t+\chi _a \D x^a )^2 +f^{-1} h_{ab} \D x^a \D x^b \,, 
\end{align}
where $f=-g_{\mu\nu}\xi^\mu \xi^\nu$ and $\chi_a$ describe respectively the 
norm and the rotation of the Killing vector. $h_{ab} $ stands for the metric 
on the quotient space of the stationary Killing field--which we shall refer to as the base space $B$--and the factor $f^{-1}$ has been inserted for convenience. 
Here and in what follows, we raise and lower the Latin indices $a, b,..$ 
by $h_{ab}$ and its inverse $h^{ab}$. 
Every metric component in (\ref{metric}) is independent of $t$. 
In this paper,  we are primarily concerned with a strictly stationary region $f>0$ unless otherwise stated.

Since $\chi_a$ describes a massless vector on the three-dimensional base space
in the sprit of Kaluza-Klein reduction, 
it can be transformed into a three-dimensional scalar $\psi$ via
\begin{align}
\label{omega}
\omega _\mu \equiv \epsilon_{\mu\nu\rho\sigma}\xi^\nu\nabla^\rho \xi^\sigma
=\nabla_\mu \psi  \,,   
\end{align}
by virtue of the identity (see e.g., \cite{wald})
\begin{align}
\label{twist}
\nabla_{[\mu} \omega_{\nu]} =-\epsilon_{\mu\nu\rho\sigma}\xi^\rho R^\sigma{}_\tau \xi^\tau \,. 
\end{align}
The system is now reduced to the three dimensional gravity coupled to the 
nonlinear sigma model  as
\begin{align}
\label{Lagrangian}
\ma L_3 = \sqrt{h } \left(R- h^{ab}G_{AB}(X)\partial_a X^A \partial_b X^B \right)\,, 
\end{align}
where $R$ is a scalar curvature built out of $h_{ab}$ and  
$X^A=(f, \psi)$. $G_{AB}$ is a metric of the target space
${\rm SU}(1,1)/{\rm U}(1)$ represented by \cite{Geroch:1970nt}
\begin{align}
\label{target}
\D s_T^2=& G_{AB}\D X^A \D X^B =
\frac{\D f^2+\D \psi^2}{2f^2 }=\frac{\D \ma E \D \bar{\ma E}}{2({\rm Re}\ma E)^2} \,,  
\end{align}
where $\ma E \equiv f- i\psi $ denotes the Ernst potential \cite{Ernst:1967wx,Ernst:1967by} 
and the bar stands for the complex conjugation. 
In terms of the stereographic coordinate
\begin{align}
\label{}
w\equiv \frac{1-\ma E}{1+\ma E} \,, 
\end{align}
the target space (\ref{target}) boils down to
\begin{align}
\label{targetw}
\D s_T^2=\frac{2\D w \D \bar w}{\Theta^2} \,, 
\end{align}
where 
\begin{align}
\label{}
 \Theta \equiv  1-w \bar w \,. 
\end{align}
The field equations are cast into 
%-------------  Lagrangian  --------------%
\begin{subequations}
\label{EOM}
\begin{align}
\label{EOMEin}
E_{ab}\equiv\, & R_{ab}-2\Theta^{-2} D_{(a} w D_{b)} \bar w=0\,, \\
\label{EOMw}
E^{(S)}\equiv\, &D^aD_a w+2\Theta ^{-1}\bar w D^a w D_aw =0\,, 
\end{align}
\end{subequations}
where $D_a$ is the derivative operator compatible with $h_{ab}$. 
One can bring 
some field equations into the divergence form~\cite{Simon}
%------------- Ernst equation  -------------%
\begin{subequations}
\begin{align}
 D^a \left(\Theta^{-1} D_a w-2i \Theta^{-2}A_a w \right)=&0\,,\\
D^a \left(\Theta ^{-2}A_a \right)=&0 \,, 
\end{align}
\end{subequations}
where 
\begin{align}
\label{Aa}
A_a \equiv  {\rm Im} (wD_a \bar w)\,.
\end{align}

The system (\ref{targetw}) is invariant under the M\"obius transformation
\begin{align}
\label{SU11}
w\to w'=\frac{\bar\alpha w+\bar\beta}{ \beta w+\alpha} \,, 
\end{align}
where $\alpha $ and $\beta$ are complex constants subjected to
\begin{align}
\label{alphabeta}
|\alpha|^2-|\beta|^2=1 \,.  
\end{align}
This ${\rm SU}(1,1)\simeq {\rm SL}(2,\mathbb R)$ isometry group of the target space (\ref{targetw}) is exploited to generate a new solution from an old one~\cite{Geroch:1970nt}. 

Since the variable $w$ transforms nonlinearly (\ref{SU11}) under ${\rm SU}(1,1)$, 
it is more propitious to work with the homogeneous coordinates $Z^A$ ($A=0, 1$) defined by 
\begin{align}
\label{}
w=\frac {Z^1}{Z^0} \,.  
\end{align}
The coordinate $w$ is regarded on the space of equivalence classes 
$Z^A\sim c Z^A$ ($c(\ne 0)\in \mathbb C$) for the 
nonzero complex vectors $Z^A$. 
This means that one can define a trivial fibre bundle $\pi: E=B\times \mathbb{CP}^2 \to B$ with a structure group ${\rm SU}(1,1)$.

It follows that the global ${\rm SU}(1,1)$ transformation acts linearly on $Z^A$ as
\begin{align}
\label{}
Z^A \to \Lambda^A {}_BZ^B \,, \qquad 
\Lambda=\left(
\begin{array}{cc}
\alpha    &    \beta \\
\bar\beta   & \bar\alpha 
\end{array}
\right)\,, 
\end{align}
where constants $\alpha $ and $\beta$  obey
(\ref{alphabeta}) for $\Lambda^A{}_B$ to be an element of 
${\rm SU}(1,1)$ as  
\begin{align}
\label{}
\Lambda^\dagger \eta \Lambda = \eta  \,, \qquad \eta_{AB} ={\rm diag}(1,-1) \,. 
\end{align}
Since one of $Z^A$ is redundant, 
one can always choose $Z^A$ to satisfy
\begin{align}
\label{ZAeq}
\langle Z, \bar Z\rangle D^aD_a Z^A= 2\Bar Z_B D^a Z^B D_a Z^a \,. 
\end{align}
where
$\bar Z_A=\eta_{AB}\bar Z^B$ and 
 $\langle Z, \bar Z\rangle =\eta_{AB}Z^A\bar Z^B$ is an ${\rm SU}(1,1) $ invariant
 norm of the complex vector $Z^A$. The three dimensional covariant derivative 
 acts on $Z^A$ trivially, i.e., $D_a Z^A=\partial_a Z^A$.

\subsection{Simon tensor}

As a complex generalization of the Cotton tensor which describes the departure from  the Schwarzschild metric in the static case,  
Simon has introduced the tensor field $S_{abc}$ as an obstruction to the Kerr solution. 
In terms of $w$, the Simon tensor reads~\cite{Simon} 
%-------------  Simon tensor   -------------%
\begin{align}
\label{Simontensor}
S_{abc}\equiv 4\Theta^{-2} (D_aD_{[b}wD_{c]}w-h_{a[b}u_{c]}) \,,
\end{align}
where 
\begin{align}
\label{}
u_a \equiv D^bD_{[b}w D_{a]}w \,,
\end{align}
satisfying $S_{abc}=S_{a[bc]} $ and $S^a{}_{ab}=0$. 
Simon has demonstrated that the Simon tensor vanishes in the asymptotically flat vacuum spacetime
if and only if the spacetime is locally isometric to the Kerr solution. 
It is a simple exercise to verify `if' part of this statement. In the Boyer-Lindquist coordinates, 
the Kerr metric reads
\begin{align}
\label{Kerr}
\D s^2 =-\frac{\Delta (r)}{\Sigma ^2 }\left(\D t-a\sin^2\theta \D \phi \right)^2
+\frac{\sin^2\theta }{\Sigma ^2 }\left((r^2+a^2)\D \phi-a \D t\right)^2
+\frac{\Sigma ^2 }{\Delta(r) }\D r^2 +\Sigma ^2 \D \theta^2\,,
\end{align}
where 
\begin{align}
 \Delta(r) =r^2-2m r+a^2, \qquad 
 \Sigma^2=r^2+a^2 \cos^2 \theta \,. 
\end{align}
The dimensional reduction along the time coordinate gives rise to the 
sigma model variables 
\begin{align}
\label{}
f= \frac{\Delta(r)-a^2 \sin^2\theta}{\Sigma^2} \,, \qquad 
\psi= -\frac{2a m \cos\theta}{\Sigma^2} \,,
\end{align}
i.e., $w=m/(r-m+ia \cos\theta)$ together 
with a base space
\begin{align}
\label{}
h_{ab}\D x^a \D x^b =\left(\Delta(r)-a^2\sin^2\theta \right)\left(
\frac{\D r^2}{\Delta(r)}+\D \theta^2
\right) +\Delta(r)\sin^2\theta \D \phi^2 \,. 
\end{align}
One can confirm that the Simon tensor (\ref{Simontensor}) vanishes for the Kerr solution.

The Cotton tensor $C_{abc}\equiv 2D_{[a}(R_{b]c}-\frac 14 R h_{b]c})$ is computed by 
equations of motion (\ref{EOM}) and  expressed in terms of the Simon tensor as 
\begin{align}
C_{[ab]c}=&\frac 12 (S_{c[ab]}+\bar S_{c[ab]})
+\frac 4{\Theta^3}D_{[a}w D_{b]}\bar w(\bar wD_c w-wD_c\bar w)\notag \\
&-\frac 2{\Theta^3}h_{c[a}(\bar wD_{b]}wD^dw-wD_{b]}\bar wD^d\bar w)D_d(w-\bar w)\notag \\
&-\frac 1{\Theta^2}[2D_cD_{[a}(w-\bar w)D_{b]}(w-\bar w)+h_{c[a}D_{b]}D_d(w-\bar w)D^d(w-\bar w)]\,.
\label{Cottonw}
\end{align}
This makes it clear that the Simon tensor $S_{abc}$ reduces to the Cotton tensor $C_{bca}$ when $w$ is real.

\subsection{Static exemplar}

The original motivation by Simon to incorporate the tensor $S_{abc}$ (\ref{Simontensor})
is to derive the divergence identities for the global characterization 
of the Kerr solution. Obviously, this hope is not attainable 
since the stationary metric (\ref{metric}) is singular on the ergosurface $f=0$
and thus fails to cover the whole domain of outer communications. 
Nevertheless, it is illustrative here to observe where this idea is sprung from.  
What has inspired Simon is the structure of field equations 
in the static spacetime\footnote{Note that $h_{ab}$ is conformally transformed by a factor
$V^{-2}$, compared with notational conventions in \cite{israel,MRS,robinson,Nozawa:2018kfk}.}
\begin{align}
\label{metricstatic}
\D s^2=-V^2 \D t^2+V^{-2} h_{ab}\D x^a \D x^b \,, 
\end{align}
where (square of) $V$ is the norm of the static Killing vector $\partial/\partial t$. 
The vacuum Einstein's equations are boiled down to 
\begin{align}
\label{eqstatic}
D^aD_a\log V=0 \,, \qquad R_{ab}=\frac 2{V^2} D_aV D_b V\,.
%D^aD_a V=0 \,, \qquad R_{ab}=\frac 1VD_aD_b V \,. 
\end{align}
The crux of the program trying to classify the static black holes 
in \cite{israel,MRS,robinson,bunting} is the Cotton tensor $C_{abc}$  for the three dimensional metric $h_{ab}$.
 With the aid of the field equations (\ref{eqstatic}), 
the Cotton tensor takes the form 
\begin{align}
\label{CottonSch}
C_{abc}=\frac{2}{V^2}\left[2D_{c}D_{[a}VD_{b]}V+h_{c[a}D_{b]}D^dV(D_dV)
-V^{-1}(DV)^2h_{c[a}D_{b]}V 
\right]\,.
\end{align}
For the Schwarzschild solution, the Cotton tensor vanishes.

In our recent paper \cite{Nozawa:2018kfk}, we have demonstrated that there exists a  meritorious  tensor field 
which can be qualified as a global characterization of the Schwarzschild 
black hole among the static solution (\ref{metricstatic}). The desired tensor is of the following form 
\begin{align}
\label{HabSch}
H_{ab} \equiv  D_a D_b V+\frac{2(1+2V^2)}{V(1-V^2)}D_aVD_bV-\frac{1+V^2}{V(1-V^2)}(DV)^2 h_{ab} \,. 
%D_a D_b V+\frac{2V}{1-V^2} \left(3D_a VD_b V-(DV)^2h_{ab} \right)\,. 
\end{align}
The traceless property $H^a{}_a=0$ follows from the equations of motion (\ref{eqstatic}). 
The previous article \cite{Nozawa:2018kfk} has given a new uniqueness proof of the 
Schwarzschild black hole among the static and asymptotically flat solutions to the vacuum
Einstein's equations. The first step of the proof consists of showing that the vanishing of $H_{ab}$ is tantamount to the Schwarzschild solution. 
This can be easily seen if we decompose the quantities on three dimensional base space
by the level set $\Sigma_V=\{V={\rm const.}\}$. The induced metric on $\Sigma_V$ is given by 
$\gamma_{ab}=h_{ab}-n_an_b$, where $n_a=\rho D_a V$ is the outward-pointing unit normal to $\Sigma_V$ 
and $\rho$ is the lapse function such that $\rho^{-2}=(DV)^2$. 
The extrinsic curvature of $\Sigma_V$ in $t=\{\rm const.\}$ surface is 
$K_{ab}=\gamma_a{}^c D_c n_b$, which is further decomposed into 
$K_{ab}=\sigma_{ab}+(1/2)K \gamma_{ab}$ where 
$\sigma_{ab}\gamma^{ab}=0$ and $K=\gamma^{ab}K_{ab}$. In terms of these 
ingredients, we have 
\begin{align}
\label{Habdec}
H_{ab}=\rho^{-1}\sigma_{ab}
-2\rho^{-2}n_{(a}\ma D_{b)}\rho 
%+\frac 1{2\rho}\left(K-\frac{4V}{\rho(1-V^2)}\right)
+\frac 1{2\rho}\left(K-\frac{2(1+V^2)}{\rho V(1-V^2)}\right)
(\gamma_{ab}-2n_a n_b) \,,
\end{align}
where $\ma D_a$ is a derivative operator built out of $\gamma_{ab}$. 
To obtain the above, we have used a relation $\partial(\log \rho)/\partial V=\rho K-1/V$
derivable from the first equation of (\ref{eqstatic}). 
Since each term on the right hand side of (\ref{Habdec}) is tensorially independent, 
$H_{ab}=0$ enforces three independent conditions
\begin{align}
\label{3cond}
\sigma_{ab}=0 \,, \qquad  \ma D_a \rho=0 \,, \qquad 
K=\frac{2(1+V^2)}{\rho V(1-V^2)}\,. 
\end{align}
Upon integration,  we immediately get the Schwarzschild solution. 
The deviation from the Schwarzschild metric is therefore encoded in $H_{ab}$. 
%$\rho =2m V/(1-V^2)^2$ and 
%$\gamma_{ab}\D x^a\D x^b=4m^2 V^2(1-V^2)^{-2} (\D \theta^2+\sin^2\theta \D \phi^2)$

The uniqueness proofs in \cite{MRS,robinson,bunting} made essential use of the 
Cotton tensor as an obstruction to the Schwarzschild solution. This is consistent with the 
aforementioned assertion, since 
the Cotton tensor (\ref{CottonSch})  is rephrased as\footnote{Note that the  tensor 
$\ti C_{ab}\equiv \frac 12 \epsilon_{acd}C^{cd}{}_b$ dual to the Cotton tensor 
is also symmetric and tracefree
$\ti C_{ab}=\ti C_{(ab)}$ and $\ti C^a{}_a=0$, but is distinct from $H_{ab}$.  
}
\begin{align}
\label{CottonHab}
 C_{abc}=\frac{2}{V^2}\left(2H_{c[a}D_{b]}V+h_{c[a}H_{b]d}D^dV  \right)\,.
\end{align}
Evidently, $H_{ab}=0$ implies $C_{abc}=0$. 
The utility of $H_{ab}$  lies in the easy handling compared to the three-index tensor $C_{abc}$. 
This becomes prominent when one tries to derive the global divergence identities. 
The three parameter family of  divergence identities constructed based upon $H_{ab}$
in  \cite{Nozawa:2018kfk} surpasses the two parameter identities in \cite{robinson}, inasmuch as the proof based on $H_{ab}$ can be readily generalized into nonvacuum case and into higher dimensions.

Last but by no means least, $C_{abc}=0$ is {\it not} equivalent to $H_{ab}=0$. 
A typical example is the topological extension of the Schwarzschild solution
$\D s^2=-f(r)\D t^2+\D r^2/f(r)+r^2\D \Sigma_k^2$, where 
$f=k-2m/r$ and $\D \Sigma_k^2$ being the metric of Einstein space with curvature $k$. 
The condition $C_{abc}=0$ is satisfied for the 
 constant time slice of this solution, but $H_{ab}=0$ is not. Since the vanishing of $H_{ab}$ requires $k=1$ with 
 $\D \Sigma_k^2$ being a metric of a unit sphere, 
$H_{ab}=0$ puts a stronger restriction on the three dimensional geometry. 
On the other hand, the condition $C_{abc}=0$ puts only the first two restrictions in (\ref{3cond}), 
which can be recognized by 
\begin{align}
\label{}
C_{abc}C^{abc}=\frac{8}{V^4\rho^4} \left(\sigma_{ab}\sigma^{ab}
+\frac{(\ma D\rho)^2}{2\rho^2}\right)\,.
\end{align}
In the conventional approach pertaining to the Cotton tensor, 
the last condition $K=2(1+V^2)/[\rho V(1-V^2)]$ must be additionally imposed 
using the rest of Einstein's equations and the resulting integration constant must be 
fixed by prescribing boundary conditions. 
By comparison to the argument  in \cite{MRS,robinson,bunting}, 
 the uniqueness proof  of \cite{Nozawa:2018kfk} based upon $H_{ab}$ thereby turns out ``one in go.''

\section{Generalization of the Simon tensor}
\label{sec:Simon}

\subsection{Alternative to the Simon tensor}
\label{sec:alternative}

In the last section, we have seen that the introduction of symmetric traceless tensor $H_{ab}$ defined by 
(\ref{HabSch}) significantly streamlines the analysis of static system, in lieu of the three-index Cotton tensor 
$C_{abc}$.  A natural question to be asked is if  a similar simplification occurs 
for the three-index complex Simon tensor (\ref{Simontensor}). We are now going to discuss that the answer is affirmative. 
We put forward  the following tracefree complex tensors as an adequate replacement of the Simon tensor: 
\begin{align}
\label{NAab}
N^A_{ab} \equiv D_a D_b Z^A-\frac 1{\langle Z, \bar Z\rangle (1-\langle Z, \bar Z\rangle)}
\bar Z_B D^c Z^B D_c Z^A h_{ab}
+\frac{1+2\langle Z, \bar Z\rangle}{\langle Z, \bar Z\rangle(1-\langle Z, \bar Z\rangle)}
\bar Z_B D_{(a}Z^B D_{b)} Z^A \,.
\end{align}
By construction, these tensors satisfy 
$N^A_{ab}=N^A_{(ab)}$ and $N^A_{ab}h^{ab}=0$, 
and linearly transform under ${\rm SU}(1,1)$ as a doublet. 
To be specific, we take a parametrization as 
\begin{align}
\label{gauge}
Z^A= \left(
\begin{array}{c}
 1   \\
w  
\end{array}
\right)\,, 
\end{align}
for which $N^0_{ab}$ vanishes trivially, whereas $N^1_{ab}(=:N_{ab}) $ is simplified to 
\begin{align}
\label{Hab}
N_{ab} = D_a D_b w+\frac{(D w)^2}{w(1-|w|^2)} h_{ab}
-\frac{3-2|w|^2}{w(1-|w|^2)}D_a w D_b w \,. 
\end{align}
A simple calculation shows that $N_{ab}=0$ is obeyed by the Kerr solution. 
One can also verify that $N_{ab}=-4V(1+V^2)^{-2} H_{ab}$ when $w=\bar w$, i.e.,  
$N_{ab}$ recovers the obstruction for the Schwarzschild metric in the static limit.  
It is noteworthy that the Simon tensor (\ref{Simontensor}) can be expressed in terms of this tensor as 
\begin{align}
\label{SabcNab}
S_{abc}=\frac{2}{ \Theta^2 }\left(2 N_{a[b}D_{c]}w + h_{a[b}N_{c]d}D^dw \right)\,.
\end{align}
This expression enjoys the  same structure as the one in the static case (\ref{CottonHab}). 
It turns out that the vanishing of $N_{ab}$ leads to $S_{abc}=0$. 
The Cotton tensor (\ref{Cottonw}) is also expressible by $N_{ab}$ more conveniently as 
\begin{align}
\label{}
C_{abc}=&\frac 1{\Theta^2} \left[2(N_{c[a}D_{b]}\bar w+\bar N_{c[a}D_{b]}w)
+h_{c[a}(N_{b]d}D^d\bar w+\bar N_{b]d}D^d w)\right] \notag \\
&+\frac {3}{|w|^2 \Theta^3}(\bar wD^dw-w D^d \bar w)h_{c[a}(D_{b]}\bar wD_d w-D_{b]}w D_d\bar w)
\notag \\ &+\frac {6}{|w|^2 \Theta^3}D_{[a}wD_{b]}\bar w (\bar w D_c w-w D_c \bar w) \,.
\label{CottonNab}
\end{align}

Several remarks are in order. 
Since $N_{ab}=0$ gives rise to nontrivial second-order differential equations for $w$, 
the existence of the solution to these equations is far from obvious beforehand.  To evince that neither inconsistency 
nor overdetermination arises, let us prospect the integrability conditions
\begin{align}
\label{}
D_{[a}N_{b]c}=&\frac{3-2|w|^2}{w(1-|w|^2)}D_{[a}w N_{b]c}
+\frac{2}{w(1-|w|^2)}D^dw N_{d[a}h_{b]c}  \notag \\
& -D_{[a }w E_{b]c}+\frac 12 E^d{}_d D_{[a}wh_{b]c}- E_{[a}{}^d h_{b]c}D_d w 
\,, 
\end{align}
and 
\begin{align}
\label{}
D^b N_{ab}=-\frac{1+2|w|^2}{ w(1-|w|^2)}N_{ab}D^bw +E_{ab}D^b w +D_a E^{(S)} 
-\frac{3-2|w|^2}{w(1-|w|^2)}E^{(S)} D_a w \,,
\end{align}
where $(E_{ab}, E^{(S)})$ have been given in (\ref{EOM}). 
It follows that the condition $N_{ab}=0$ does not conflict with Einstein's equations of motion
$E_{ab}=E^{(S)}=0$. 

By the reasoning identical to the static case, the conditions imposed by 
$S_{abc}=0$ are more relaxed than $N_{ab}=0$. The Kerr metric satisfies both, but its topological generalization fails to fulfill $N_{ab}=0$.  In this light, the condition $N_{ab}=0$ gives rise to an unequivocal characterization of the Kerr solution. 

It is also worthwhile to stress that we have fixed the gauge as $Z^A=(1,w)$ to derive the explicit expression of $N_{ab}$ in (\ref{Hab}). 
We caution the reader that we are not allowed to perform a further ${\rm SU}(1, 1)$ transformation by merely replacing $w$ in $N_{ab}$ as (\ref{SU11}), since $N_{ab}^A$ transforms as a fundamental representation of ${\rm SU}(1,1)$.  As a matter of fact, the explicit form of $N^2_{ab}$ 
nontrivially changes and $N^1_{ab}$ becomes no longer identically trivial under (\ref{SU11}), for which 
both of the components $N^A_{ab}$ play the role of obstruction for the Kerr-NUT family. 
The only allowed transformation that leaves the form of (\ref{Hab}) invariant
is  $Z^A=(1,w) \mapsto (e^{i\theta}, e^{-i\theta}w)$ with $\theta \in \mathbb R$, 
which corresponds to the maximal subgroup ${\rm U}(1)$ of ${\rm SU}(1,1)$. 
On the other hand, the Simon tensor $S_{abc}$ is inert 
under ${\rm SU}(1, 1)$, since equation (\ref{SabcNab}) can be recast into 
an ${\rm SU}(1, 1)$ singlet as
\begin{align}
\label{}
S_{abc}=\frac{2}{\langle Z,\bar Z\rangle^2} \epsilon_{AB}\epsilon_{CD}Z^A Z^C \left(
2N^B_{a[b}D_{c]}Z^D+h_{a[b}N^B_{c]d}D^d Z^D 
\right) \,,
\end{align}
where $\epsilon_{AB}$ is an alternating tensor. 

Bearing this gauge fixing in mind, 
we are now going to derive a number of equations of the divergence type by making use of $N_{ab}$. 
To this aim, we notice the following identities
\begin{align}
\label{}
D^2 \rho=& -\rho^3 (D_aD_bw)^2 +\frac 3\rho (D \rho)^2-\rho^3 D^a w[D_a(D^2w)+R_{ab}D^bw ]
\notag \\
=& -\rho^3 (D_aD_bw)^2  +\frac 3\rho (D \rho)^2+\frac{2\bar w^2}{\rho\Theta^2} -\frac{4\bar w}
\Theta D_a\rho D^a w \,, 
\end{align}
and
\begin{align}
\label{}
(D_aD_b w)^2 =N_{ab}N^{ab} +\frac{2(3-2|w|^2)}{w\Theta} N_{ab} D^a w D^b w
+\frac{1}{\rho^4 w^2\Theta^2}(6-8|w|^2+4|w|^4) \,, 
\end{align}
where we have supposed
\begin{align}
\label{rho}
\rho \equiv (D^a w D_a w)^{-1/2} 
\end{align}
is well-defined. 
Squaring (\ref{SabcNab}) yields 
\begin{align}
\label{}
S_{abc}S^{abc}=\frac{8}{\rho^2\Theta^4} \left(N_{ab}N^{ab}-\frac 3{2\rho^2} N_aN^a\right)\,, 
\end{align}
where 
\begin{align}
\label{Na}
N_a \equiv  \frac{D_a \rho }{\rho }+\frac{2D_a w}{w} =-\rho^2 N_{ab}D^b w \,. 
\end{align}
Capitalizing on these relations, 
one obtains the following divergence equations\footnote{It should be noted that the second term on the right hand side of (\ref{Simonsq2}) is missing in eq. (22) of \cite{Simon}. }
\begin{align}
\label{Simonsq1}
D^a [\Theta ^{-1}D_a k -2i\Theta^{-2}k A_a ] &=\frac 1{16}k^{-7}
 \Theta^3 S_{abc}S^{abc} \,, \\
 \label{Simonsq2}
D^a [\Theta^{-1}(w D_a k -k D_a w)] & =\frac 1{16}w k^{-7}\Theta^3 S_{abc}S^{abc}
+\frac{kw}{2 \Theta^2} N_a (\bar w D^a w-w D^a \bar w) \,. 
\end{align}
where $A_a$ has been defined by (\ref{Aa}) and 
\begin{align}
\label{}
k\equiv (D^a w D_a w)^{1/4}=\rho ^{-1/2}\,, 
\end{align}
has been introduced to conform to the notation in \cite{Simon}. 
In addition to those constructed in \cite{Simon}, 
we obtain the following one parameter family of divergence equation
\begin{align}
 \label{Simonsq3}
D_a \left[(\rho w^2)^c \left(\frac{|w|^2}{\Theta}N^a{-}\frac{2i A^a}{c\Theta^2}\right)\right]
=-\frac{|w|^2  (\rho w^2)^c}{\Theta}\rho^2\left(N_{ab}N^{ab}-\frac{2+c}{\rho^2}N_aN^a \right) \,, 
\end{align}
where $c$ is an arbitrary real constant. 

As opposed to the static case, there are two main obstacles when one tries to 
apply these equations to characterize the four dimensional metric. The first difficulty is that the 
metric (\ref{metric}) only covers the outside region of the ergosurface $f=0$. The stationary metric form is not well adapted to the global boundary value problems. 
Another adversity is that the right hand side of these equations (\ref{Simonsq1})--(\ref{Simonsq3})
remain complex. To the contrary, the corresponding expressions in the static case
can be made positive-definite, so that these divergence equations are 
utilized for proving the uniqueness  for global boundary value problems. It therefore seems no direct applicability of 
these equations (\ref{Simonsq1})--(\ref{Simonsq3}) in Lorentzian signature. 
Nevertheless, these equations would be of great help for the uniqueness proof of 
gravitational instanton solutions in Euclidean signature \cite{Simon:1995ty}.

\subsection{Taxonomy of solutions with $N_{ab}=0$}

This section provides the local form of the four dimensional stationary metric admitting $N_{ab}=0$. 
The original argument of the metric reconstruction for $S_{abc}=0$ is given by Perjes \cite{Perjes}. 
Since the condition $N_{ab}=0$ is stronger than $S_{abc}=0$, 
our proof allows one to obtain the explicit metric form and discloses the underlying geometric structure more manifest.

The classification divides into subcases according to 
(I) $D_aw$ is not null and (II) $D_aw$ is null. Case (I) is further 
categorized into (I-i) $\D w \we \D \bar w \ne 0$ and 
(I-ii) $\D w \we \D \bar w=0$.

\bigskip\noindent{\bf Case (I-i).} In this case, the variable $w$ and its complex conjugation 
can be exploited for the coordinates in the base space. 
One can then define the following real vector field
\begin{align}
\label{Ua}
U^a =\frac{i (1-|w|^2)}{(w\bar w)^3}\epsilon^{abc}D_bw D_c \bar w \,, 
\end{align}
which satisfies $U^a D_a w=0$, i.e., $U^a$ is orthogonal to $D_a w$ and $D_a \bar w$. 
From $N_{ab}=0$, it can be straightforwardly checked  that
\begin{align}
\label{}
D_{(a}U_{b)} =0 \,, \qquad 
U_{[a}D_b U_{c]} =0 \,. 
\end{align}
Namely, 
$U^a$ defines a hypersurface-orthogonal Killing vector of the base space. 
It is also advantageous to adopt the following symmetric tensor 
\begin{align}
\label{KT}
K_{ab}\equiv \frac{1-w\bar w}{w^3\bar w^3}\left((D^cw)(D_c\bar w)h_{ab}-D_{(a}wD_{b)}\bar w\right) \,.
\end{align}
When $N_{ab}=0$, this tensor satisfies
\begin{subequations}
\begin{align}
\label{}
D_{(a}K_{bc)}& =0\,, \\
\mas L_U K_{ab}&=0\,.
\end{align}
\end{subequations}
It therefore turns out that $K_{ab}$ defines the Killing tensor of the base space, which is invariant under the action of the Killing vector $U^a$. 
The Killing tensor (\ref{KT}) is irreducible in the sense that it is neither proportional to the metric nor 
expressed as a symmetric product of Killing vectors. 
This class of space is considerably restrained and its canonical metric falls into the 
Benenti-Francaviglia form \cite{Benenti}. 
Without adopting $N_{ab}$, it would have been impossible to identify 
this unexpected hidden symmetry. This represents one of the advantages of using $N_{ab}$ 
relative to $S_{abc}$.

Let us move on to introduce  the local coordinate system admitting $N_{ab}=0$. Since $U^a$ is a Killing vector, 
it is opportune to employ the coordinate ($w, \bar w, \varphi$) in 
such a way that $U^a =(\partial/\partial \varphi)^a$.
Specifically, the inverse metric $h^{ab}$ can be put into  
\begin{align}
\label{metricinv}
h^{ab}\partial_a \partial_b 
= U^{-2} \partial_\varphi^2 +\rho^{-2} \partial_w^2 +\bar \rho^{-2} \partial_{\bar w}^2
+\frac{2\Omega }{\rho \bar \rho}\partial_w \partial_{\bar w} \,,
\end{align}
where ($U, \Omega$) are undetermined real functions of ($w, \bar w$) and 
$h^{ww}=(Dw)^2=\rho^{-2}$ comes from the definition (\ref{rho}). 
An inspection of (\ref{Na}) immediately yields
\begin{align}
\label{}
\rho =\frac{C}{w^2} \,,  
\end{align}
where $C$ is a complex constant. 
Computing the norm of 
(\ref{Ua}), one finds 
\begin{align}
\label{Usol}
U(w,\bar w)=\frac{1-|w|^2}{|C|^2 |w|^2}\sqrt{\Omega ^2-1} \,.  
\end{align}
By $N_{ab}D^b \bar w=0$ and $N_{ab}U^b=0$, $\Omega$ is determined to be 
\begin{align}
\label{betasol}
\Omega(w, \bar w)=\frac{1}{2|w|^2 (1-|w|^2)} \left(
\frac{\bar C}{C}w^2+\frac{C}{\bar C}\bar w^2 +2C_0w^2\bar w^2 
\right)\,,
\end{align}
where $C_0$ is a real constant. 

We have exhausted all the constraints coming from $N_{ab}=0$. 
One verifies that all the Einstein's equations of motion (\ref{EOM}) are satisfied. 
To cast the metric into a more recognizable form, let us change 
 $(w, \bar w)$ to real coordinates $ (q, p)$ by
$w=C/(q+i p)$. 
Denoting 
\begin{align}
\label{}
C=m-i n \,, \qquad (m,n )\in \mathbb R\,, 
\end{align} 
let us further shift $q\to q-m$ and $p\to p+n$, i.e., 
\begin{align}
\label{}
w=\frac{m-in }{(q-m)+i (p+n)} \,.
\end{align}
Introducing a new real constant $a$ by 
\begin{align}
\label{}
C_0= -1 +\frac{2a^2}{m^2+n^2} \,, 
\end{align}
we find 
\begin{align}
\label{3Dbase}
h_{ab}\D x^a \D x^b =\big(Q(q)-P(p)\big)\left(\frac{\D q^2}{Q(q)}+\frac{\D p^2}{P(p)}\right)+Q(q)P(p) \D \sigma^2 \,, 
\end{align}
where 
\begin{align}
\label{}
Q(q)\equiv q^2-2mq+a^2 -n^2 \,, \qquad 
P(p) \equiv a^2-(n+p)^2 \,,
\end{align}
and 
\begin{align}
\label{}
\varphi=\frac 12 (m^2+n^2)^2 \sigma \,.   
\end{align}
The sigma model variables are 
\begin{align}
\label{}
f=\frac{Q(q)-P(p)}{q^2+p^2}\,, \qquad \psi=-\frac{2(mp+nq)}{q^2+p^2}\,.
\end{align}
By means of these coordinates, the Killing tensor (\ref{KT}) is 
\begin{align}
\label{3DKT}
K_{ab}\D x^a \D x^b =\frac{1}{(m^2+n^2)^2}\left[
(Q(q)-P(p)) \left(\frac{Q(q)}{P(p)}\D p^2+\frac{P(p)}{Q(q)}\D q^2\right)
+(Q(q)+P(p))Q(q)P(p)\D \sigma^2 
\right] \,.
\end{align}
This Killing tensor is responsible for the separability of the 
geodesic motion $k^bD_b k^a=0$ for the three dimensional space (\ref{3Dbase}).

Letting $\tau $ be a stationary coordinate, 
the four dimensional metric then reads
\begin{align}
\label{Carter}
\D s^2=-\frac{Q(q)}{q^2+p^2}(\D \tau-p^2 \D \sigma ) ^2+\frac{P(p)}{q^2+p^2}(\D \tau+q^2 \D \sigma)^2 +(q^2+p^2)\left(\frac{\D q^2}{Q(q)}+\frac{\D p^2}{P(p)}\right)\,. 
\end{align}
This is nothing but the Carter form \cite{Carter:1968ks} of the Kerr-NUT solution, where 
$m$, $a$, $n$ denote respectively the mass, the specific angular momentum and the NUT charge. 
The physical interpretation of the NUT charge has been given from various perspectives in
\cite{Turakulov:2001bm,Bossard:2008sw,Argurio:2008zt,Mukherjee:2018dmm}.
The Kerr metric (\ref{Kerr}) is recovered by 
\begin{align}
\label{}
q=r \,, \qquad p=a \cos\theta \,, \qquad n=0  \,, \qquad \tau=t-a \phi \,, \qquad \sigma=-\frac{\phi}{a} \,. 
\end{align}
It appears that the three dimensional Killing tensor (\ref{3DKT}) presented above is not related to the eminent four dimensional Killing tensor~\cite{Carter:1968ks,Walker:1970un}.

\bigskip\noindent{\bf Case (I-ii).} 
Letting $w=w_{\rm R}+i w_{\rm I}$ ($w_{\rm R}, w_{\rm I}\in \mathbb R$), 
the condition $\D w \we \D \bar w=0$ implies that $w_{\rm I}$ is a function of $w_{\rm R}$. 
Equivalently, $f=f(\psi)$ expressed in terms of the original variable $w=[1-(f-i\psi)]/[1+(f-i \psi)]$. 
This class of solutions is referred to as a Papapetrou's class \cite{Papapetrou}
(see also section 20.3 in \cite{Stephani:2003tm}). 
Since $\psi=0$ is exhausted by the Schwarzschild solution, 
we postulate $\D \psi \ne 0$ in the sequel. 
Inserting $f=f(\psi)$ into equations of motion (\ref{EOMw}) of $w$, 
we thus get 
\begin{align}
\label{eqPapapetrou}
\psi=C_2+\sqrt{C_1-f^2} \,, \qquad 
D^a D_a \left({\rm arctanh}\frac{\sqrt{C_1-f^2}}{\sqrt C_1}\right)=0\,, 
\end{align}
where $C_1$ and $C_2$ are real constants. 
Equation (\ref{CottonNab}) implies that the base space $h_{ab}$ is conformally flat. Note that 
this conclusion is not drawn immediately from  (\ref{Cottonw}). 
The conformal flatness permits one to 
foliate the base space by the level set $f={\rm const.}$ as \cite{MRS,robinson,bunting}
\begin{align}
\label{}
h_{ab}\D x^a \D x^b =F_1 ^2 (f)\D f^2 +F_2^2 (f) \gamma_{ij}(y^k)\D y^i \D y^j\,,
\end{align}
where $\gamma_{ij}$ is a two-dimensional metric on the $f={\rm const.}$ surface and can be 
taken to be a conformally flat form $\gamma_{ij}(y^k)\D y^i \D y^j=e^{2\Phi}\D z \D \bar z$. 
The second equation in (\ref{eqPapapetrou}) is integrated to give
\begin{align}
\label{}
F_2(f)= \left(C_3 f(C_1-f^2)^{1/2}F_1(f)\right)^{1/2} \,, 
\end{align}
where $C_3$ is a constant. Substitution of this into  Einstein's 
equations $E_{ab}D^a f D^b f=0$ yields
\begin{align}
\label{}
F_1 (f)=C_5 \frac{f+(\sqrt{C_1}-f)C_1C_4^2+\sqrt{C_1}(1+2C_4\sqrt{C_1-f^2})}
{\sqrt{C_1-f^2}(\sqrt{C_1}+f-(\sqrt{C_1}-f)C_1C_4^2)^2}\,. 
\end{align}
where $C_4$ and $C_5$ are integration constants. 
Plugging this into $N_{ab}=0$, one finds 
\begin{align}
\label{}
C_2=-\sqrt{C_1-1}\,, \qquad C_4=\frac{\sqrt{C_1} +1}{\sqrt{C_1(C_1-1)} } \,.
\end{align}
The rest of Einstein's equations is satisfied, provided that  
the conformal factor in $\gamma_{ij}(y^k)\D y^i \D y^j=e^{2\Phi}\D z \D \bar z$
satisfies the Liouville equation
\begin{align}
\label{}
\partial_z \partial_{\bar z} \Phi =-\frac 14 k e^{2\Phi} \,, \qquad
k\equiv \frac{C_1C_3}{(\sqrt{C_1}-1) C_5} \,. 
\end{align}
The local solution to this equation can be chosen to be
\begin{align}
\label{}
 e^{\Phi}= \frac 1{1+k z \bar z/4} \,,
\end{align}
for which $\D \Sigma_{k}^2 \equiv e^{2\Phi}\D z \D \bar z$ is a space of constant curvature $k$, 
which can be normalized to be $k=\pm 1, 0$. 
Define a new variable $r$ by $r=\int F_1(f)\D f$, i.e., 
\begin{align}
\label{}
f=\frac{r^2-n^2-2m r}{r^2+n^2}\,, \qquad 
\psi=\frac{2n(r-m)}{r^2+n^2}\,, 
\end{align}
where we have renamed the constants 
$C_3=2kn$ and $C_1=(m^2+n^2)/n^2$. 
In the region $r^2-2mr-n^2>0$, $F_2$ becomes real for $k=1$. 
Changing to $z=2\tan(\theta/2)e^{i\phi}$, 
the base space therefore admits ${\rm SO}(3)\simeq {\rm SU}(2)$ symmetry
\begin{align}
\label{}
h_{ab}\D x^a \D x^b =\D r^2+(r^2+n^2)f(r) (\D \theta^2+\sin^2\theta \D \phi^2)\,. 
\end{align}
The rotation of the Killing vector reads $\chi=2n \cos\theta \D \phi$. 
It follows that the four dimensional metric gives rise to the Taub-NUT solution
\cite{Taub:1950ez,Newman:1963yy}.

\bigskip\noindent{\bf Case (II).} 
Also in this case, the vector $U^a$ defined in (\ref{Ua}) is well-defined. 
For $D^a w D_a w=0$, 
this vector satisfies $D_a U_b=0$, i.e., $U_a$ is a Killing vector that is covariantly constant. 
One can then cast the metric into 
\begin{align}
\label{}
h_{ab}\D x^a \D x^b =2  f_1(w, \bar w)\D w \D \bar w+\D v^2 \,, 
\end{align}
where $U^a=(\partial/\partial v)^a$ and $f_1$ is a real function. 
The condition $N_{ab}=0$ is immediately integrated as
\begin{align}
\label{}
f_1=\frac{D_1^2(1-|w|^2)}{ w^3 \bar w^3}\,,
\end{align}
where $D_1$ is a real constant. Equations of motion (\ref{EOM}) 
demand no more restrictions.
Introducing new coordinates ($r, \theta$) by $w^{-1}=r e^{i\theta}$, 
the base space simplifies to
\begin{align}
\label{}
h_{ab}\D x^a \D x^b=2D_1^2(r^2-1)(\D r^2+r^2 \D \theta^2)+\D v^2 \,.  
\end{align}
The twist one-form $\chi=\chi_a\D x^a$ of the Killing vector can be chosen to 
\begin{align}
\label{}
\chi= \frac{2(1+r \cos\theta)}{r^2-1}\D v\,. 
\end{align}
The vacuum four dimensional metric (\ref{metric})
takes a more suggestive form by $t=u+v$ as
\begin{align}
\D s^2=&-\frac{r^2-1}{r^2+1+2r \cos\theta}\D u^2-2\D v\D u
+2D_1^2(r^2+1+2r \cos\theta)(\D r^2+r^2 \D \theta^2) \notag\\
=& -\left(1-\frac{\sqrt{D_1}}{\sqrt{2\zeta}}-\frac{\sqrt{D_1}}{\sqrt{2\bar\zeta}}\right)\D u^2
-2\D v\D u+2 \D \zeta \D \bar \zeta \,, 
\label{ppwave}
\end{align}
where $\zeta =D_1(1+re^{i\theta})^2/2$. 
When promoted into four dimensions, the vector $\partial/\partial v$ 
is null and covariantly constant. This class of solutions is 
referred to as a $pp$-wave \cite{Stephani:2003tm}.

\subsection{Four dimensional description}

During the course of our analyses so far, we have limited the discussion for the Simon tensor and its generalization within the framework of the quotient space associated with the stationary Killing field. This formulation inevitably encounters some obstacles as described at the end of 
section \ref{sec:alternative}. To partially overcome this shortcoming,  
Mars has provided a four dimensional counterpart of the Simon tensor (\ref{Simontensor})
in \cite{Mars:1999yn}.   

The Mars tensor $M_{\mu\nu\rho}$ is defined as
\begin{align}
\label{Marstensor}
M_{\mu\nu\rho}\equiv 4E^+_{\mu[\nu}\sigma_{\rho]} -\frac 12 h_{\mu[\nu} \xi^\lambda C^+_{|\lambda|\rho]\sigma\tau}(\D \xi^+)^{\sigma\tau} \,,
\end{align}
where 
\begin{align}
\label{}
E^+_{\mu\nu}\equiv C^+_{\mu\rho\nu\sigma}\xi^\rho \xi^\sigma \,, \qquad 
\sigma_\mu \equiv\xi^\nu (\D \xi^+)_{\nu\mu}\,, 
\end{align}
and
\begin{subequations}
\begin{align}
\label{}
C^+_{\mu\nu\rho\sigma} =&\, C_{\mu\nu\rho\sigma} +\frac i2 \epsilon_{\mu\nu\tau\lambda}
C^{\tau\lambda}{}_{\rho\sigma} \,, \\
(\D \xi^+)_{\mu\nu} =&\, 2\nabla_{[\mu}\xi_{\nu]} +i  \epsilon_{\mu\nu\rho\sigma}
\nabla^{\rho}\xi^\sigma \,, \\
h_{\mu\nu}= &\, fg_{\mu\nu}+ \xi_\mu \xi_\nu \,.  
\end{align}
\end{subequations}
Imposing the vacuum Einstein's equations, the vector $\sigma_\mu$ 
can be represented by the Ernst potential 
as $\sigma_\mu =\nabla_\mu (f-i\psi)=\nabla_\mu \ma E$. 
Owing to $h_{\mu\nu}\xi^\nu=0$, $h_\mu{}^\nu$ can be viewed as a projection operator 
orthogonal to $\xi^\mu$ and $h_{ab}$ is identified as an induced metric on the base space. 
The Mars tensor satisfies $M_{\mu\nu\rho}=M_{\mu[\nu\rho]}$, $M^{\mu}{}_{\mu\nu}=0$ and 
$M_{\mu\nu\rho}\xi^\rho=0=M_{\mu\nu\rho}\xi^\mu$, so that the only nonvanishing components 
are obtained by projecting onto the base space. 
When $R_{\mu\nu}=0$, the relation (\ref{twist}) implies 
\begin{align}
\label{}
\nabla_\mu (\D \xi^+)_{\nu\rho}=2C^+_{\rho\nu\mu\sigma}\xi^\sigma \,.
\end{align}
Together with $(\D \xi^+)_{\mu\nu}(\D \xi^+)^{\mu\nu}=-(4/f)\sigma_\mu \sigma^\mu$, 
the Mars tensor reduces to 
\begin{align}
\label{}
M_{\mu\nu\rho}=2 \left(\nabla_\mu \sigma_{[\nu }-\nabla_\mu \xi^\sigma (\D \xi^+)_{\sigma[\nu} \right)\sigma_{\rho]} 
+\frac 12 h_{\mu[\nu}\nabla_{\rho]} (f^{-1}\sigma_\tau\sigma^\tau) \,. 
\end{align}
To simplify the second term in the above equation, 
one resorts to 
\begin{align}
\label{}
\nabla_\mu \xi_\nu=-\frac 12 f^{-1}\epsilon_{\mu\nu\rho\sigma}\xi^\rho\omega^\sigma
-f^{-1}\xi_{[\mu} \nabla_{\nu]}f\,, \qquad 
(\D \xi^+)_{\mu\nu}=-\frac 1f(2\xi_{[\mu}\sigma_{\nu]}+i\epsilon_{\mu\nu\rho\sigma}\xi^\rho \sigma^\sigma)\,, 
\end{align}
where $\omega_\mu$ is defined by (\ref{omega}),  
leading to 
\begin{align}
\label{}
\nabla_\mu \xi^\sigma (\D \xi^+)_{\sigma\nu} =\frac 1{2f}\nabla_\mu f \sigma_\nu-\frac 1{2f^2}\xi_\mu\xi_\nu
\sigma^\rho \nabla_\rho f +\frac i{2f}(\omega_\nu\sigma_\mu -f^{-1} h_{\mu\nu} \omega_\rho \sigma^\rho)
-\frac 1{f^2}\xi_{(\mu}\epsilon_{\nu) \rho \sigma\tau}\nabla^\rho f \xi^\sigma \omega^\tau \,. 
\end{align}
By projecting onto the base space guided by the method in \cite{Geroch:1970nt}, some calculations show that 
\begin{align}
\label{}
M_{abc}= &
2 D_a \sigma_{[b}\sigma_{c]} 
+\sigma^d h_{a[b}\left(D_{c]}\sigma_{d}-\frac 1f\sigma_{c]}\sigma_d\right)=
\frac{2(1-|w|^2)^2}{(1+w)^4}S_{abc} \,. 
\end{align}
This demonstrates the identification of these tensors up to the multiplicative factor. 
As contrasted to the fact that the Simon tensor (\ref{Simontensor}) is defined only outside
the ergosphere,  the Mars tensor (\ref{Marstensor}) does not suffer from this difficulty. 
Furthermore, the Mars tensor is defined irrespective of the existence of Ernst potentials.

Inspecting the form of the Mars tensor (\ref{Marstensor}), it is tempting to inquire the spacetime picture of the tensor (\ref{Hab}). For this purpose, the following tensor is 
a well-suited candidate of this kind 
\begin{align}
\label{maN}
 \ma N_{\mu\nu}=E^+_{\mu\nu}+\frac 1{2\sigma^2} E^+_{\rho\sigma}\sigma^\rho \sigma^\sigma
 \left(f^{-1} h_{\mu\nu}-\frac{3\sigma_\mu\sigma_\nu}{\sigma^2}\right)\,,
\end{align}
where $\sigma^2\equiv \sigma_\mu \sigma^\mu\ne 0$ has been assumed. 
By a straightforward computation, it can be verified that $ \ma N_{\mu\nu}=0$ is satisfied by the 
Kerr-NUT metric (\ref{Carter}). This statement can be slightly strengthened in such a way that 
$ \ma N_{\mu\nu}=0$ is indeed satisfied for the metric (\ref{Carter}) with {\it arbitrary} structure functions ($Q(q), P(p)$), i.e., insensitive to the satisfaction of Einstein's equations. 
The above tensor obeys $\ma N_{\mu\nu}=\ma N_{(\mu\nu)}$, $\ma N_\mu{}^\mu=0$ and $\ma N_{\mu\nu}\xi^\nu=0$, so that $\ma N_{ab}$ is identified as a tensor on the base space.

To demonstrate that the  on-shell value of $\ma N_{\mu\nu}$ recovers (\ref{Hab}) when projected onto the base space, it is useful to record
\begin{align}
\label{}
E^+_{ab}=\frac 12 D_aD_b \ma E+\frac 1{4f} D_a \ma E D_b \ma E-\frac 1{4f}D^c \ma E
D_c \ma E h_{ab} \,.
\end{align}
We then find
\begin{align}
\label{Nabrel}
\ma N_{ab}=-\frac 14 (1+\ma E)^2 \left[
N_{ab}+\frac 1{2(D\ma E)^2} N_{cd}D^c \ma E D^d \ma E \left(h_{ab}-3\frac{D_a \ma E D_b \ma E}{(D \ma E)^2}\right)
\right] \,.
\end{align}
Accordingly, the on-shell  $\ma N_{\mu\nu}$ stores the geometric data identical to $N_{ab}$, 
as we wanted to show. 

By noting 
$\xi^\lambda C^+_{\lambda\mu \sigma\tau}(\D \xi^+)^{\sigma\tau}=-(4/f)E^+_{\mu\nu}\sigma^\nu$, 
one can rewrite the Mars tensor (\ref{Marstensor}) into 
\begin{align}
\label{}
M_{\mu\nu\rho}=2 \left(2\ma N_{\mu[\nu}\sigma_{\rho]} + f^{-1} h_{\mu[\nu}\ma N_{\rho]\sigma}\sigma^\sigma \right) \,.
\end{align}
This equation retains a striking pattern reminiscent of (\ref{CottonHab}) and (\ref{SabcNab}). 

It follows from these parallel structures that equation (\ref{Nabrel}) defines the 
spacetime characterization of (\ref{Hab}) that admits a description simpler than the Mars tensor. 
However, this should not be taken too literally. 
The expression in (\ref{maN}) incorporates the inverse of $f=-g_{\mu\nu}\xi^\mu \xi^\nu$  and 
also $\sigma^2=f h^{ab}D_a\ma E D_b\ma E$, both of which are not well-defined at the ergosurface. 
Unfortunately, the global property that has been a major meliority of the Mars tensor is now lost. 
This is a fundamental limitation of our tensor (\ref{maN}).

Nevertheless, the use of $\ma N_{\mu\nu}$ rather than $M_{\mu\nu\rho}$ unveils a new 
geometric condition for the Kerr-NUT family in the following way. 
By construction, we have
\begin{align}
\label{Cartercond}
\ma N_{\mu\nu}\sigma^\nu =E^+_{\mu\nu}\sigma^\nu-\frac{E^+_{\rho\sigma}\sigma^\rho \sigma^\sigma}{\sigma^2}\sigma_\mu =0\,. 
\end{align}
This is an eigenvalue problem which is well-defined insensitive to the sign of $f$, provided 
$\sigma^2 \ne 0$.\footnote{For the $pp$-wave metric (\ref{ppwave}), we have $\sigma^2=0$ and $E^+_{\mu\nu}\sigma^\nu=0$. } 
It follows that the configuration $\ma N_{\mu\nu}=0$ implies that 
$\sigma^\mu$ is an eigenvector of $E^+_{\mu\nu}$. 
This is the four dimensional covariant condition free from restriction $f>0$ and symbolizes the (off-shell) Kerr-NUT family (\ref{Carter}). This simple criterion is appealing and has been 
 unnoticed in the literature. Note that the diagonalizability of $E^+_{\mu\nu}$ with doubly degenerate eigenvalues amounts to the Petrov D condition. 
 As a consistency check, the condition (\ref{Cartercond}) is not satisfied by 
the most general Petrov-D vacuum metric constructed by Pleba\'nski and Demia\'nski~\cite{Plebanski:1976gy}.

\section{Electrovacuum}
\label{sec:Maxwell}

Let us now extend the discussion of the vacuum case in the previous section 
into the electrovacuum. 
The field equations to the Einstein-Maxwell system read
\begin{align}
\label{}
R_{\mu\nu}=2 \left(F_\mu{}^\rho F_{\nu\rho}-\frac 14 g_{\mu\nu}F_{\rho \sigma}F^{\rho\sigma}\right)  \,, \qquad \D F=\D \star F=0 \,. 
\end{align}
Assuming that the Maxwell field is also invariant under the flow of the stationary Killing 
vector $\mas L_\xi F=0$, the Maxwell equation and the Bianchi identity 
imply the existence of local scalar functions ($E, B$) such that 
\begin{align}
\label{}
\nabla _\mu E=F_{\mu\nu}\xi^\nu \,, \qquad 
\nabla_\mu B=-\star F_{\mu\nu}\xi^\nu \,. 
\end{align}
The twist $\omega_\mu =\epsilon_{\mu\nu\rho\sigma}\xi^\nu \nabla^\rho\xi^\sigma $
of the Killing vector satisfies (\ref{twist}), leading to
\begin{align}
\label{}
\nabla_{[\mu }\omega_{\nu]}=4 \nabla_{[\mu}E\nabla_{\nu]} B \,.
\end{align}
This implies the local existence of a twist scalar $\psi $ such that 
\begin{align}
\label{}
\omega_\mu = \nabla_\mu \psi +2 (E\nabla_\mu B -B \nabla_\mu E )\,. 
\end{align}
Then, the system is reduced to the gravity-coupled sigma model (\ref{Lagrangian})
with the target space ${\rm SU}(1,2)/{\rm S}({\rm U}(1)\times {\rm U}(1,1))$~\cite{Mazur:1983vi,Eris:1984tu}
\begin{align}
\D s_T^2 =&\, \frac 1{2f^2}\big[\D f^2+(\D \psi+2E\D B-2B \D E)^2\big]
-\frac{2}{f} (\D E^2+\D B^2) \notag \\
=&\, \frac 1{2({\rm Re} \ma E+|\Phi|^2)^2} (\D \ma E+2\bar \Phi \D \Phi )(\D \ma{\bar E}+2\Phi \D \bar\Phi )
-\frac{2}{{\rm Re} \ma E+|\Phi|^2}\D \Phi \D \bar \Phi \,,
\label{target0}
\end{align}
where
\begin{align}
\label{}
\ma E\equiv f -i\psi -(E^2+B^2) \,, \qquad 
\Phi \equiv -E+i B \,. 
\end{align}
For the spacelike reduction $f<0$, 
this is a negative curvature complex projective space 
$\mathbb{CP}^{1,1}$, which admits the simultaneous K\"ahler and quaternionic structures. 
In the stationary reduction $f>0$, the target space (\ref{target0}) has the ($+,+,-,-$) signature. 

By the following holomorphic transformation to the new complex variables $w^i$ ($i=1,2 $) 
\begin{align}
\label{}
w^1=\frac{1-\ma E}{1+\ma E} \,, \qquad w^2=\frac{2\Phi}{1+\ma E } \,, 
\end{align}
the field equations reduce to 
\begin{subequations}
\label{EMeq}
\begin{align}
\label{EMeq1}
R_{ab}=&\, 2\Theta^{-1} D_{(a}w^i D_{b)} \bar w_i+2\Theta^{-2}w_i\bar w_j D_{(a}\bar w^i D_{b)}w^j \,, \\
\label{EMeq2} D^aD_a w^i=&  -2\Theta^{-1} \bar w_j  D^a w^iD_a w^j  \,, 
\end{align}
\end{subequations}
where 
\begin{align}
\label{}
w_i =\eta_{ij}w^j \,, \qquad 
\eta_{ij}={\rm diag}(1,-1) \,,
\end{align}
and 
\begin{align}
\label{}
\Theta\equiv 1-  w\cdot \bar w \,, \qquad 
 v_1\cdot \bar v_2 \equiv 
\eta_{ij}v_1^i \bar v_2^j \,. 
\end{align}
One can introduce the complex vectors $Z^A$ ($A=0,1,2$)  by~\cite{Kinnersley:1973,Mazur:1983vi}
\begin{align}
\label{}
w^1=\frac{Z^1}{Z^0}\,, \qquad 
w^2=\frac{Z^2}{Z^0}\,.
\end{align}
Apart from the distinction that the indices $A, B$ now run from 
0 to 2 with
\begin{align}
\label{}
\eta_{AB}={\rm diag}(1,-1,1)\,, 
\end{align}
these quantities $Z^A$ can be chosen to satisfy (\ref{ZAeq}), 
where $\langle Z, \bar Z\rangle =\eta_{AB}Z^A\bar Z^B$ and 
$\bar Z_A=\eta_{AB}\bar Z^B$ as before.

The generalization of the complex tensor $N^A_{ab}$ into the Einstein-Maxwell system is 
straightforward and is given by the same form as (\ref{NAab}): 
\begin{align}
\label{NAab:EM}
N^A_{ab} \equiv D_a D_b Z^A-\frac 1{\langle Z, \bar Z\rangle (1-\langle Z, \bar Z\rangle)}
\bar Z_B D^c Z^B D_c Z^A h_{ab}
+\frac{1+2\langle Z, \bar Z\rangle}{\langle Z, \bar Z\rangle(1-\langle Z, \bar Z\rangle)}
\bar Z_B D_{(a}Z^B D_{b)} Z^A\,, 
\end{align}
 where $Z^A$ is now regarded as a triplet of ${\rm SU}(1,2)$.
 Besides this, the obstruction must supplemented by
\begin{align}
\label{Nab:EM}
\mas N_{ab} \equiv \epsilon_{ABC} Z^A D_a Z^B D_b Z^C  \,. 
\end{align} 
Here $\epsilon_{ABC}$ is an alternate tensor of  ${\rm SU}(1,2)$. 
We affirm that both of $N^A_{ab} $ and $\mas N_{ab} $ 
fulfill a role as an obstruction to the Kerr-Newman-NUT family. 
Because of the formal similarity, 
the most of the discussion in the vacuum case carries over to the electrovacuum case. 

 Choosing the homogeneous coordinates as $Z^A=(1, w^i)$, these tensors are  boiled down to
\begin{align}
\label{Niab}
N^{i}{}_{ab} \equiv  D_a D_b w^i+\frac 1{ w\cdot \bar w \Theta}\bar w_k D^c w^k D_c w^i h_{ab}-\frac{3-2w\cdot \bar w }{w\cdot \bar w  \Theta} \bar w_k D_{(a} w^k D_{b)} w^i \,, 
\end{align}
and 
\begin{align}
\label{Nab:EM2}
\mas N_{ab}=2D_{[a} w ^1 D_{b]} w^2 \,. 
\end{align}
This symmetric tensor satisfies $N^{i}{}_a{}^a=0$ in view of the Einstein-Maxwell field equations
(\ref{EMeq}). 
When the electromagnetic field is switched off $w^2=0$, one recovers $N^{1}{}_{ab}\to N_{ab}$ in (\ref{Hab}). 
We stress that the form of the tensors (\ref{Niab}) and (\ref{Nab:EM2}) remains unchanged for
\begin{align}
\label{stabilitysub}
Z^A =\left(\begin{array}{c}
1    \\
w^1\\
w^2   
\end{array}\right) \mapsto \Lambda^A{}_BZ^B\,, \qquad 
\Lambda= \left(
\begin{array}{ccc}
e^{2i\theta} & 0 & 0    \\
 0 &   e^{-i\theta}\alpha & e^{-i\theta}\beta \\
 0 & e^{-i\theta}\bar \beta & e^{-i\theta} \bar\alpha 
\end{array}
\right) \,, \qquad |\alpha^2|-|\beta|^2=1\,, 
\end{align}
corresponding to the stability subgroup ${\rm S}({\rm U}(1)\times{\rm U}(1,1))$
of ${\rm SU}(1,2)$.

Let us now construct solutions admitting $N^{i}{}_{ab}=\mas N_{ab}=0$. 
The condition (\ref{Nab:EM2}) implies that $w^2$ is a function only of $w^1$, viz, $w^2=h(w^1)$. 
Inserting this into the equations of motion (\ref{EMeq2}), one arrives at $(Dw^1)^2\partial ^2h/\partial (w^1)^2=0$. 
For concreteness, we shall assume that that $D_aw^1$ is not null or zero. 
Then, this leads to 
\begin{align}
\label{}
w^2=\gamma w^1 +\delta \,, 
\end{align}
where $\gamma $ and $\delta $ are complex constants. 
We denote $w^1=:w$ and $w^2=\gamma w+\delta $ in what follows.

From $N^1_{ab}D^b w=0$, one finds a relation $D_a (\log \rho) \propto D_a w$ for 
$\rho^{-2}=(Dw)^2$. The integrability 
$D_{[a}D_{b]}\log \rho=0$ of this equation is assured only for (i) $w$ is real or (ii) $\delta =0$. 
For real $w$ with $\delta\ne 0$, one can foliate $h_{ab}\D x^a \D x^b =\rho^2(w)\D w^2 +\Xi(w)^2 e^{2\Phi(z,\bar z)}\D z\D \bar z$, for which 
one finds no solutions compatible with Einstein's equations. It follows that the only allowed possibility is (ii) $\delta =0$. 

The rest of the proof is fairly straightforward. 
For the sake of clarity, we focus on the case $\D w \we \D \bar w \ne 0$. 
The vanishing of $N^1{}_{ab}$ implies that the following vector is a hypersurface-orthogonal Killing vector
\begin{align}
\label{}
U^a = \frac{i [1-(1-|\gamma|^2)|w|^2]}{(w\bar w)^3}\epsilon^{abc}D_b w D_c \bar w \,.  
\end{align}
By $N_{ab}^1D^b w=0$, we have $\rho =C w^{-2}$ with $C$ being a complex constant.
For the metric ansatz (\ref{metricinv}), we thus obtain
\begin{align}
\label{}
U=\frac{1-(1-|\gamma|^2)|w|^2}{|C|^2|w|^2}\sqrt{\Omega^2-1} \,, \qquad 
\Omega= \frac{1}{2|w|^2[1-(1-|\gamma|^2)|w|^2]} \left(\frac{\bar C}{C}w^2+\frac{C}{\bar C}\bar w^2 
+2C_0 w^2\bar w^2\right)\,. 
\end{align}
Choosing 
\begin{align}
\label{}
C=m-in \,, \qquad 
C_0=-1+\frac{2a^2+Q_e^2+Q_m^2}{m^2+n^2} \,, \qquad \gamma=-\frac{Q_e+i Q_m}{m-i n} \,,
\end{align}
and transforming to $w=(m-in)/[q-m+i(p+n)]$ with $\phi=\frac 12 (m^2+n^2)^2\sigma$, 
we get the base space (\ref{3Dbase}) with structure functions
\begin{align}
\label{}
Q(q)=q^2-2m q+a^2-n^2+Q_e^2+Q_m^2 \,, \qquad 
P(p)=a^2-(n+p)^2 \,.
\end{align}
This recovers the Kerr-Newman-NUT family \cite{DemianskiNewman} with the electric charge $Q_e$ and 
the magnetic charge $Q_m$.

\section{Summary}
\label{sec:summary}

This paper has endeavored to give a coherent description for the local characterization of the Kerr-NUT family \cite{Simon}. 
We have proposed a symmetric traceless complex tensor (\ref{NAab}) or (\ref{Hab}) defined on the 
orbit space of the stationary Killing vector. This tensor takes the place of the Simon tensor in several respects. 
Our tensor (\ref{NAab}) is an ${\rm SU}(1,1)$ vector-valued symmetric tensor and is 
simplified  in the gauge (\ref{gauge}) to the form (\ref{Hab}), by means of which the Simon tensor 
is entirely described as (\ref{SabcNab}). 
Most notably, the practical benefit of our tensor (\ref{Hab}) is its simple tensorial structure, which allows us to  elucidate the hidden symmetry (\ref{KT}) on top of the Killing symmetry (\ref{Ua}). 
Apart from the degenerate cases which correspond to 
Taub-NUT and $pp$-waves, our proposed tensor is entitled as a new criterion for the 
local identification of the Kerr-NUT family. Moreover, the obstruction tensor is readily generalized into the electrovacuum case, while leaving the ${\rm SU}(1,2)$ invariance unbroken. 
This is the main importance of the present result. 

We have further undertaken to covariantize the  tensor (\ref{Hab}) in the language of four dimensions, following
the philosophy of Mars \cite{Mars:1999yn}. 
Since our expression (\ref{maN}) is not globally defined, it does not seem to have a significant advantage over the 
Mars tensor. Despite this unsatisfactory feature, we have found a new criterion (\ref{Cartercond}) as a four dimensional covariant obstruction to the Kerr-NUT metric, which is much more manageable and fully amendable to analytic study. 

Several applications of our results are conceivable. 
One can consider various relatives of the Kerr-NUT metric.  
The Wahlquist class of solutions \cite{Wahlquist:1968zz} describing a rigidly rotating perfect fluid in 
stationary and axisymmetric family is one of the nonvacuum extensions of the Carter solution, since 
it admits a Killing-Yano tensor with a three-form torsion \cite{Hinoue:2014zta}. 
One can verify that the Wahlquist metric satisfies $\ma N_{\mu\nu}=0$ (and $M_{\mu\nu\rho}=0$
\cite{Marklund:1996zy}) only on-shell. 
This traces back to the fact that the Wahlquist metric belongs to type I for off-shell, whereas
to type D for on-shell \cite{Houri:2019nun}. It would be an interesting future direction to explore the 
relation to the (torsionful) Killing-Yano tensor and the tensor $\ma N_{\mu\nu}$ and to the quotient space 
interpretation. 

Throughout the paper, we have addressed the local characterization of the Kerr-NUT solution. 
For the application of the global boundary value problems for the stationary and axisymmetric system, 
we need to perform the first dimensional reduction  along the Killing vector that generates ${\rm U}(1)$
\cite{Carter:1971zc,Robinson:1975bv,Mazur:1982db} instead of the stationary Killing vector.  
To date, no adequate counterpart of the Simon tensor has been constructed for the spacelike reduction. 
It would be definitely worthwhile to pursue this direction for the new uniqueness proof of rotating black holes. 
Work along this direction is in progress.

An intrinsic labeling of the Kerr-NUT metric has been also given in \cite{Ferrando:2008nw}, 
where no stationary assumption is made. Alternatively, additional conditions should be 
presumed therein. They gave two criteria: (i) the eigen-twoform of  the complex selfdual 
Weyl tensor in Petrov-D space should satisfy a certain differential equation, 
(ii) the gradients of the Weyl scalar invariants for Petrov-D space should satisfy
a certain algebraic equation. 
 The analysis of \cite{Ferrando:2008nw} follows in part the spirit of the Cartan-Karlhede program, 
according to which the equivalence problem and the isometry group of a given metric can be 
addressed by the Riemann tensor and its covariant derivatives. Recently, this Cartan-Karlhede program
has been streamlined substantially into a practical form by \cite{Kruglikov:2018qcn,Nozawa:2019dwu,Nozawa:2019pff} in three dimensions. 
It is then a promising route  to examine generalizations of these algorithms into $3+1$ dimensions, 
for the labeling the Kerr-NUT solution under more relaxed conditions.

The method developed here seems to be operative only in four spacetime dimensions. 
The Myers-Perry metric~\cite{Myers:1986un} possibly with a cosmological constant and a NUT charge can be written into the Carter form in arbitrary dimensions~\cite{Chen:2006xh}. It would be interesting to explore the 
analogue of the Simon tensor for the dimensionally reduced space of these higher dimensional metrics.

\acknowledgments
The present work is partially supported by 
Grant-in-Aid for Scientific Research (A) from JSPS 17H01091 and (C) 20K03929.

%%%%%%%%%%%%%%%%%%%%%%%%%%%%%%%
%                                                                                                %
%                                   References                                         %
%                                                                                                %
%%%%%%%%%%%%%%%%%%%%%%%%%%%%%%%

\end{document}